\documentclass[twocolumn,superscriptaddress,pra]{revtex4}
\usepackage{graphicx}
\usepackage{amsmath}
\usepackage{amsthm}
\usepackage{amssymb}
\usepackage{latexsym}
\usepackage{array}
\usepackage{float}
\usepackage{amsfonts}
\usepackage{mathrsfs}
\usepackage{verbatim}

\usepackage{color}

\begin{document}

\title{Experimental demonstration of quantum learning speed-up with classical input data}

\author{Joong-Sung~Lee}
\thanks{The first three authors contributed equally to this work}
\affiliation{Department of Physics, Hanyang University, Seoul 04763, Korea}

\author{Jeongho~Bang}
\thanks{The first three authors contributed equally to this work}
\affiliation{School of Computational Sciences, Korea Institute for Advanced Study, Seoul 02455, Korea}
\affiliation{Institute of Theoretical Physics and Astrophysics, University of Gda\'{n}sk, 80-952 Gda\'{n}sk, Poland}

\author{Sunghyuk~Hong}	
\thanks{The first three authors contributed equally to this work}
\affiliation{Department of Physics, Hanyang University, Seoul 04763, Korea}

\author{Changhyoup~Lee}
\affiliation{Institute of Theoretical Solid State Physics, Karlsruhe Institute of Technology, 76131 Karlsruhe, Germany}

\author{Kang Hee~Seol}
\affiliation{Department of Physics, Hanyang University, Seoul 04763, Korea}

\author{Jinhyoung~Lee}
\thanks{Corresponding authors}
\affiliation{Department of Physics, Hanyang University, Seoul 04763, Korea}

\author{Kwang-Geol~Lee}
\thanks{Corresponding authors}
\affiliation{Department of Physics, Hanyang University, Seoul 04763, Korea}

\received{\today}

\begin{abstract}
We consider quantum-classical hybrid machine learning in which large-scale input channels remain classical and small-scale working channels process quantum operations conditioned on classical input data. This does not require the conversion of classical (big) data to a quantum superposed state, in contrast to recently developed approaches for quantum machine learning. We performed optical experiments to illustrate a single-bit universal machine, which can be extended to a large-bit circuit for binary classification task. Our experimental machine exhibits quantum learning speed-up of approximately $36\%$, as compared to the fully classical machine. In addition, it features strong robustness against dephasing noise.

\end{abstract}

\maketitle

\newcommand{\bra}[1]{\left<#1\right|}
\newcommand{\ket}[1]{\left|#1\right>}
\newcommand{\abs}[1]{\left|#1\right|}
\newcommand{\expt}[1]{\left<#1\right>}
\newcommand{\braket}[2]{\left<{#1}|{#2}\right>}
\newcommand{\commt}[2]{\left[{#1},{#2}\right]}

\newcommand{\tr}[1]{\mbox{Tr}{#1}}

\newcommand{\note}[1]{\textcolor{blue}{#1}}
\newcommand{\issue}[1]{\textcolor{red}{#1}}

\section{Introduction}

Quantum machine learning (QML) has recently attracted significant interests~\cite{Wittek14,Schuld15,Adcock15,Biamonte17,Lau17,Monras17}. However, this approach is confronted with several challenges and questions, which include: is there any advantage in quantum machine learning compared to its classical counterpart? If so, what are the quantum effects and how do they contribute to this advantage? These issues were addressed theoretically in terms of learning performance, such as membership query complexity~\cite{Servedio04,Kothari13} and sample complexity~\cite{Arunachalam16} (see Ref.~\cite{Ciliberto17} for the relevant summary). Various quantum learning scenarios have also been proposed, e.g., quantum neural networks~\cite{Silva16,Wan17}, quantum Boltzmann machine~\cite{Amin16}, and quantum reinforcement learning~\cite{Dunjko16,Dunjko14}.

Recent progress has been made in several specific learning applications such as data classification~\cite{Yoo14,Rebentrost14}, regression~\cite{Zhao15,Schuld16}, topological data analysis~\cite{Lloyd16}, and anomaly detection~\cite{Liu18}. One of the noteworthy areas of research is quantum support vector machine, which is a method to classify big quantum data~\cite{Rebentrost14,Cai15,Li15}. The quantum support vector machine and its variants exploit the quantum algorithm for solving linear equations (often called HHL, named after the inventors, Harrow, Hassidim, and Lloyd~\cite{Harrow09}), to facilitate an exponential speed-up of quantum learning. These machines operate with the limits of restrictive stipulations~\cite{Aaronson15}. For example, the HHL algorithm requires the kernel matrix to be sparse. The condition number must also scale sub-linearly with the system size. Furthermore, the machines need to be equipped with a quantum random access memory~\cite{Giovannetti08-1,Giovannetti08-2}, where big input data are rapidly converted and efficiently resourced to allow for quantum superpositioning. It is challenging to realize quantum random access memory without encountering a few well-known problems. For example, the impracticality of error correction, the requirement of uniformly distributed data over a quantum register, and the intrinsic bound of memory latency~\cite{Arunachalam15,Aaronson15}. We are thus interested in another QML
method; namely, one that does not involve the conversion of (big) classical input data into quantum-superposed information~\cite{Yoo14}.

In classical machine learning, a machine learner receives a finite set of sampled inputs $\mathbf{x}=x_1 x_2 \cdots x_N$ ($x_j \in \{0, 1\}$ for all $j=1,\ldots,N$) and their labels $c(\mathbf{x})$. Here, $c$ is a target function that the learner is supposed to learn, which is called a concept. For the given training data $\{\mathbf{x}, c(\mathbf{x}) \}$ and a set of hypotheses $h$, classical machine learning is formally defined by the identification of a hypothesis $h$ which is close to the concept $c$~\cite{Langley96}. In contrast, the QML process begins with the preparation of training data in quantum states, e.g., $\{\ket{\mathbf{x}}, \ket{c(\mathbf{x})} \}$, and the processing of superposed states using quantum operations. However, in this instance, we consider another approach in which the input training data are not converted to a quantum state, but the operations are processed based on a quantum mechanical approach. We refer to this method as the quantum-classical hybrid scheme. This approach will be beneficial from the perspective of quantum random access memory.

In this report, we consider a universal machine of binary classification with input and output, each consisting of a single classical bit and an internal working channel of a qubit, based on a hybrid approach. This is a basic machine, with classical input channels that can be extended to an arbitrary number of bits, while its internal working channel is maintained with a single qubit~\cite{Dunjko18}. We performed optical experiments to illustrate the single-bit universal machine functionality for a binary classification. Our experimental machine exhibits a quantum advantage of approximately $36\%$ as compared to a fully classical machine. The system also exhibits strong robustness against dephasing.

\section{Preliminaries}

\subsection{Binary classification} 

A simple and representative example of binary classification is the operation of an email-filter, which classifies incoming emails as spam and non-spam categories by referring to the labels of the emails. A single bit y is introduced for each label which indicates whether an email is spam, so that $y=0$ indicates it is a spam and otherwise, it is not. An $N$-bit string $\mathbf{x} = x_1 x_2 \cdots x_N$ is also introduced to denote the features of an email, e.g., name, address, date, etc. The machine is supposed to learn a map from $\mathbf{x}$ to $y$ so that the machine eventually classifies the emails. Candidates for such a map are $N$-bit Boolean functions, 
\begin{eqnarray}
\mathbf{x} \in \{0, 1\}^{N} \to y \in \{0, 1\}.
\label{eq:Booleanf}
\end{eqnarray}
The candidate maps are called hypotheses in the language of machine learning. The goal of machine learning in this scenario is to find a hypothesis $h$ that is acceptable with a small error $E$ as for the email filter~\cite{Opper91,Langley96}. 

A circuit can be introduced for realizing the entire set of hypotheses in Eq.~(\ref{eq:Booleanf}). This circuit consists of an $N$-bit input channel to handle the input data $\mathbf{x}$ and the one-bit working channel that processes operations to realize a hypothesis. The working channel runs $2^{N}$ controllable gate operations: one single-bit operation and $2^{N}-1$ operations conditioned on the bit values $x_j$ of $\mathbf{x} = x_1 x_2 \cdots x_N$. These gate operations are supposed to be either {\small identity} (doing nothing) or logical-not (flipping a bit). This circuit is universal since it realizes all possible ($2^{2^N}$) Boolean functions~\cite{Yoo14,Gupta06}. 

It is worth mentioning that as $N$ is increased, it will be impractical to require $2^N$ gates for the circuit, and approximate answers with even smaller gates will suffice for practical applications. The latter can be made by excluding unlikely hypotheses based on {\em a priori} information~\cite{Chapelle09}.

\subsection{Universal single-bit feature circuit} 

\begin{figure}
\centering
\includegraphics[width=0.46\textwidth]{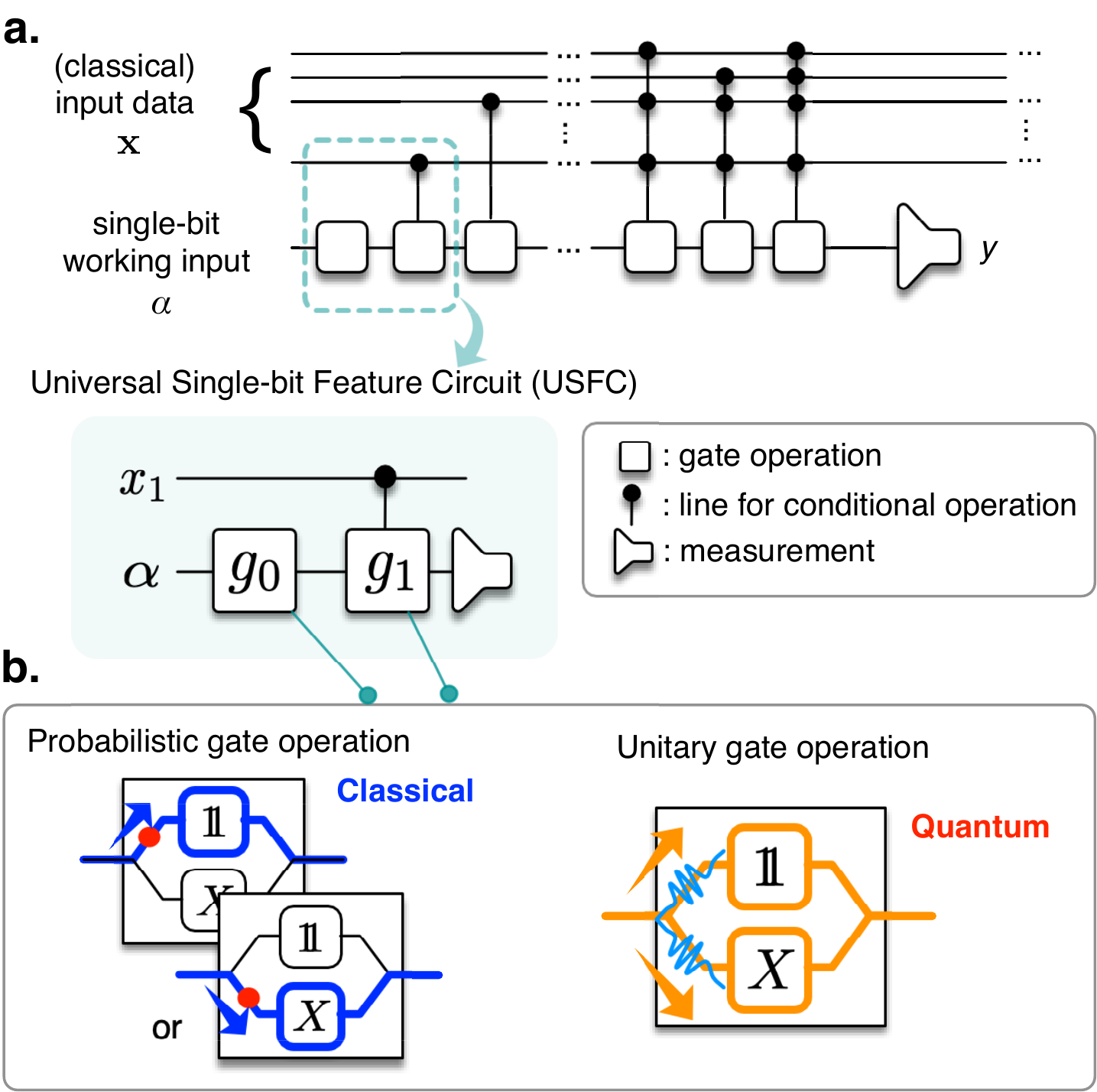}
\caption{\label{fig:circuit} {\bf The binary classification circuits.} {\bf a}. We consider a reversible circuit that consists of two types of channels: one deals with $N$-bit classical input data and the other provides a working bit. This circuit runs $2^N$ gate operations to realize all possible $N$-bit Boolean functions: one single-bit operation and $2^{N}-1$ operations conditioned by the bits $x_j$ of $\mathbf{x}$. These gate operations are supposed to be either identity ($\openone$) or logical-not ($X$). Note that such a circuit can realize all possible ($2^{2^N}$) binary classification functions~\cite{Gupta06}. We focus on the basic machine of single-bit circuit, referred to as the universal single-feature circuit. {\bf b}. Here, we consider two versions of the basic circuits: one is the classical circuit that operates the probabilistic operations, and the other is the quantum circuit that operates the unitary operations utilizing quantum interference.}
\end{figure}

We shall focus on a basic universal machine throughout the paper, which we call a universal single-bit feature circuit (USFC). The USFC receives a single-bit feature $x$ from the input channel and operates two gates $g_{0,1}$ in the internal working channel (see Fig.~\ref{fig:circuit}{\bf a}). The USFC thus realizes four possible deterministic tasks: [$\boldsymbol{\tau.1}$] $x \to y=0$, [$\boldsymbol{\tau.2}$] $x \to y=x$, [$\boldsymbol{\tau.3}$] $x \to y=1$, and [$\boldsymbol{\tau.4}$] $x \to y=x \oplus 1$. The USFC is crucial for its generalization, noting that the latter exhibits a quantum speed-up that is more than quadratic~\cite{Yoo14}, that extends the classical input channels to an arbitrary number of bits and maintains the internal working channel with a single qubit. In addition, the analysis of USFC can provide some intuition to speed-up learning (see Appendix~A). Since the generalization is made by increasing the classical bits and it is experimentally feasible, the realization of the USFC is central to this work.

USFC learning is performed as described in this section for a given target task, which is represented by one of [$\boldsymbol\tau.1$]-[$\boldsymbol\tau.4$] if it is deterministic or by a conditional probability $\mathrm{Pr}_\tau(y|x)$ if it is probabilistic. Input data that are transmitted through the classical channels remaining unaltered. They condition the operations to be performed in the working channel. The initial signal in the working channel is successively flipped as it passes through the two gates $g_k$ ($k=0,1$) with $g_1$ conditional on the input bit. A measurement is performed at the end of the working channel. The measurement outcome is used to evaluate how close the current circuit is to the target $c$, and the gates $g_k$ ($k=0,1$) are adjusted based on the evaluation in the interim. Here, the learning control of $g_k$ is assumed to be made based on the probability, which we refer to as the gate-adopting preference,
\begin{eqnarray}
\text{Pr}(g_k \to \openone) ~~\text{and}~~ \text{Pr}(g_k \to X) = 1 - \text{Pr}(g_k \to \openone),
\label{eq:probs}
\end{eqnarray}
which are the probabilities that the gate $g_{k}$ operates the identity $\openone$ and the logical-not $X$, respectively. Such probabilistic operations can offer practical advantages in a heuristic manner~\cite{Langley96,Ghahramani15}.

We introduce the classical and quantum machines of USFC for comparison. Noting that the input channels are classical for both types of machines~\cite{Yoo14}, by the classical USFC (cUSFC) we mean the classical realization of the working channel; the working input $\alpha$ is a binary number that is initially defined as $\alpha = 0$, and the gates $g_k$ ($k=0,1$) run randomly ($\openone$ or $X$) with probabilities given in Eq.~(\ref{eq:probs}). The operations $g_k$ ($k=0,1$) are represented by a stochastic evolution matrix in the basis of classical bits:
\begin{eqnarray}
\left(
\begin{array}{cc}
\text{Pr}(g_k \to \openone) & \text{Pr}(g_k \to X) \\
\text{Pr}(g_k \to X) & \text{Pr}(g_k \to \openone)
\end{array}
\right),
\end{eqnarray}
where the gate-adopting preferences are written as the transition probabilities. On the other hand, the quantum USFC (qUSFC) works with the quantum working channel and unitary gate operations. The channel is assumed to be initially in a quantum state $\ket{\alpha} = \ket{0}$, where $\{\ket{0}, \ket{1}\}$ is the computational basis of a qubit. The gate operations are represented in the computational basis as:
\begin{eqnarray}
\left(
\begin{array}{cc}
\sqrt{\text{Pr}(g_k \to \openone)} & e^{i \phi_k}\sqrt{\text{Pr}(g_k \to X)} \\
e^{-i \phi_k}\sqrt{\text{Pr}(g_k \to X)} & -\sqrt{\text{Pr}(g_k \to \openone)}
\end{array}
\right),
\end{eqnarray}
which involves the intrinsic probabilistic nature. Thus, both types of cUSFC and qUSFC are treated on equal footing, disregarding the quantum phases. In this way, we single out the quantum phases $\phi_{0,1}$ and their roles in machine learning, and expect that they are engaged in quantum interference between two operations (as depicted in Fig.~\ref{fig:circuit}{\bf b})~\cite{Yoo14}.

\subsection{Quantum learning speed-up} 

In this subsection, we briefly review the results in Ref.~\cite{Yoo14}. The error $E=1-F$ of the learning is analyzed with the task fidelity F that indicates how close a circuit performs the desired task. The task fidelity F is defined by:
\begin{eqnarray}
F=\left( \prod_{\mathbf{x}}\sum_y \sqrt{\text{Pr}(y|\mathbf{x}) \text{Pr}_\tau (y|\mathbf{x})} \right)^{1/{2^N}}, 
\label{eq:fidelity}
\end{eqnarray}
where $\text{Pr}(y|\mathbf{x})$ is the conditional probability that the machine learner produces $y$ for given $\mathbf{x}$ at a certain stage of learning, and $\text{Pr}_\tau(y|\mathbf{x})$ are the target probabilities that a given task determines, as in Refs.~\cite{Yoo14,Cover12}. Then we obtain the difference of the quantum and classical task fidelities for $N=1$ as
\begin{eqnarray}
F_\text{Q}^4 - F_\text{C}^4 = \Lambda \cos{\Delta},
\label{eq:F_cq}
\end{eqnarray}
where 
\begin{eqnarray}
\Lambda &=& 2 \text{Pr}(g_0 \to \openone) \sqrt{\prod_{k=0,1}\text{Pr}(g_k \to \openone) \text{Pr}(g_k \to X)}, \nonumber \\
\Delta &=& \abs{\phi_1 - \phi_0}. 
\end{eqnarray}
The subscripts C and Q denote classical and quantum, respectively. Here, Eq.~(\ref{eq:F_cq}) is obtained for [$\boldsymbol{\tau.1}$] with $\{ \text{Pr}_\tau(0|0)=1, \text{Pr}_\tau(0|1)=1 \}$, while similar results are obtained for the other tasks [$\boldsymbol{\tau.2}$]-[$\boldsymbol{\tau.4}$]~\cite{Yoo14}. 

The principal implication of Eq.~(\ref{eq:F_cq}) is the expansion of regions for the approximate hypotheses to the given target task by appropriately choosing the quantum phases $\phi_{0,1}$. A wider region of the approximate hypotheses implies that it is easier for the machine to find one of them in the entire hypothesis space. Here, the approximate hypotheses for the target task are defined by the condition $E = 1-F \le \epsilon_t$ where $\epsilon_t$ is an error tolerance. Note that these implications originate from quantum superpositioning~\cite{Yoo14}. It is thus crucial to appropriately choose the quantum phases $\phi_{0,1}$ for the speed-up of qUSFC learning. One could offset the advantage $\Lambda$ in Eq.~(\ref{eq:F_cq}) with $\cos{\Delta}=0$ and even transform it to a disadvantage with $\cos{\Delta}=-1$. Similar behavior is found for the different phases of the tasks [$\boldsymbol{\tau.2}$]-[$\boldsymbol{\tau.4}$]. The quantum learning speed-up is more pronounced with an $N$-bit circuit by using the coherence of all $2^N$ unitary gates (see Ref.~\cite{Yoo14} for more details).

\section{Experiment with Heralded Single Photons}

\subsection{Linear-optical settings for USFC learning} 

\begin{figure}
\centering
\includegraphics[width=0.46\textwidth]{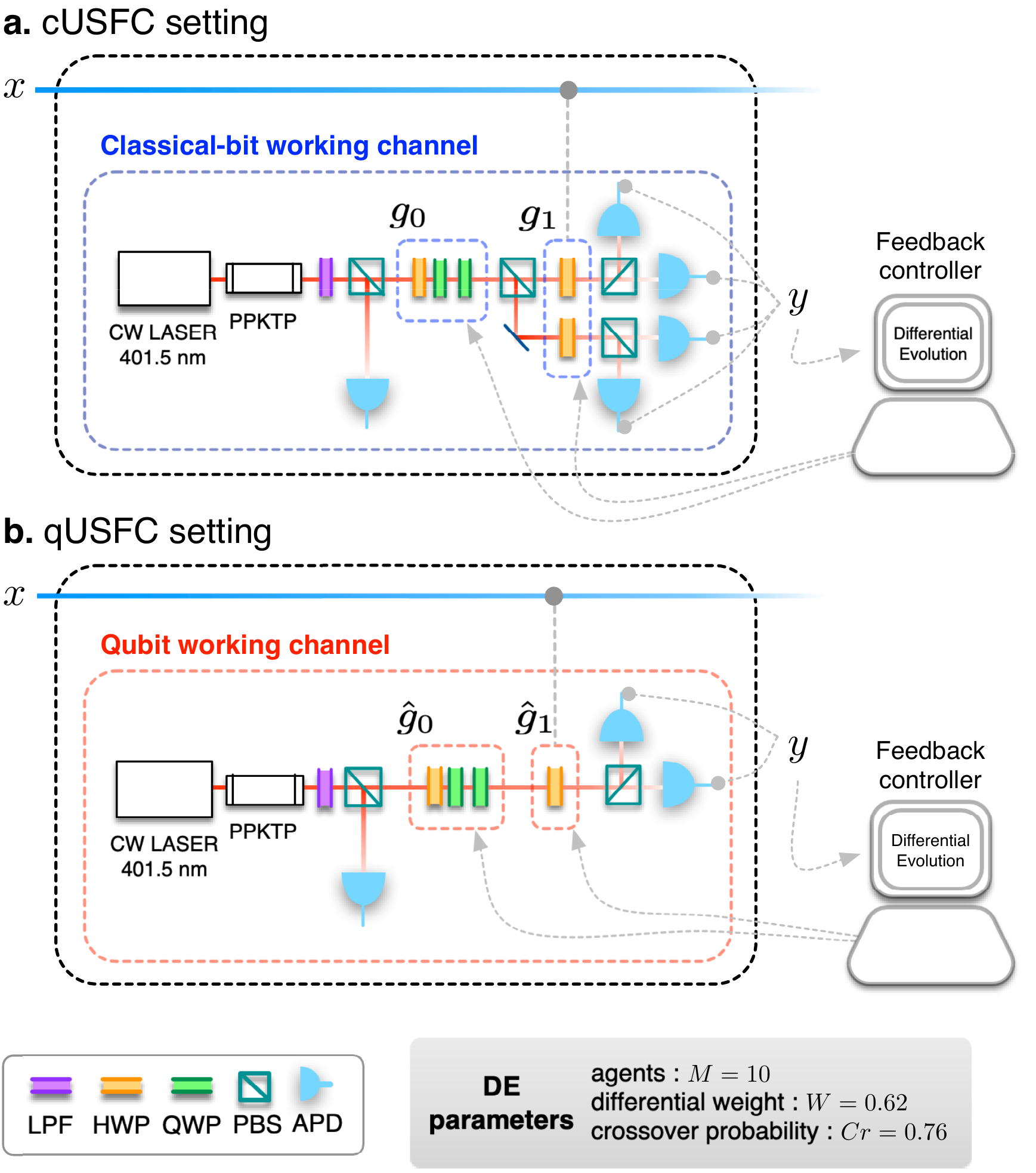}
\caption{\label{fig:setting} {\bf Schematic view of our experimental setups.} We design two versions of USFC learning experiments (i.e., {\bf a} classical and {\bf b} quantum) in the linear-optical regime, where single-photon polarizations, i.e., horizontal ($H$) and vertical ($V$), are used as single-bit information carriers of the working channel. In quantum USFC learning, the quantum superposition effects involved in single-photon polarization are exploited. On the other hand, in classical USFC learning, the quantum superposition of photon polarization is completely destroyed when passing through the polarization beam splitter (PBS) placed between $g_0$ and $g_1$. In our experiments, differential evolution (DE)~\cite{Storn97} is employed as a learning algorithm.}
\end{figure}

Given the classical input data $x \in \{0,1\}$, the USFC can be realized in a linear-optical platform with single photons as the working bit $\alpha$, as shown in Fig.~\ref{fig:setting}. The orthogonal polarization of a single photon in free space constitute the basis for the bit; horizontally ($H$) or vertically ($V$) polarized single photon represents the bit $0$ or $1$ in the working channel. In our experiment, the working input $\alpha=0$ (i.e., $H$-polarized single photon) is prepared by a heralded single-photon generation scheme. We pumped a periodically poled $\text{KTiOPO}_4$ (PPKTP) crystal of a length $10$~mm with a continuous wave pump laser at a wavelength $401.5$~nm (MDL-III-400, CNI). Pairs of orthogonally polarized photons at $799.2$~nm and $803.5$~nm with a FWHM of $6.7$~nm and $5.0$~nm are produced in the same spatial mode via phase-matching for collinear type-II spontaneous parametric down-conversion. The photon pairs are divided using a polarization beam splitter (PBS) placed after the $750$~nm long pass filter (LPF). The reflected $V$-polarized photon is directly measured in the idler mode by an avalanche photodiode (APD, SPCM-AQR-15, PerkinElmer), whereas the transmitted $H$-polarized photon is sent to the working channel of the USFC circuit and measured by the APDs at the end of the circuit. We post-select only the cases when coincidence detection occurs between the idler and working channel after synchronization of the arrival times of the two optical paths. This allows us to exclude the cases where the photon in the working channel is lost, e.g., loss in the optics and detectors. Coincidence detection in our experiment is analyzed using a field-programmable gate array (FPGA, PXI-7841R, National Instruments) with a clock speed of $40$~MHz. 

The gate operations $g_0$ and $g_1$ are expected to vary the polarization of a single photon in the working channel. This is achieved in a controlled manner by choosing appropriate angles for the optic axis of birefringent crystals, e.g., half or quarter wave plates (HWP or QWP), through which the initial $H$-polarized photon is transmitted. Specifically, we implement the gate operation $g_0$ using a stack of crystals, HWP($\vartheta_0$)-QWP($\varphi_0$)-QWP(${\pi}/{4}$). Here, $\vartheta_0$ and $\varphi_0$ denote the angles between the horizontal axis and the fast axes of the first two wave-plates to be controlled, and the last one is fixed with an angle of ${\pi}/{4}$. The gate $g_1$, on the other hand, is realized using a single plate, HWP($\vartheta_1$) with $\vartheta_1$. These hands-on rotation angles handle the gate-adopting preferences such that~\cite{Damask04}:
\begin{eqnarray}
\text{Pr}(g_0 \to \openone)&=&\cos^{2}\left(2\vartheta_{0}-\varphi_{0}-\frac{\pi}{4}\right), \nonumber \\ 
\text{Pr}(g_1 \to \openone) &=&\cos^{2}2\vartheta_{1}, \nonumber \\
\Delta&=&2\varphi_{0}+\frac{\pi}{2}.
\end{eqnarray}
The USFC is also equipped with a feedback controller, which is responsible for the learning; the feedback controller updates the angles with no {\em a priori} knowledge.

Using such a setting, the USFC is expected to identify a set of (say, optimal) angles so that it eventually becomes a realization of the desired task. To this end, the change in polarization is measured at the end, and a number of the identical measurement for single-photon inputs are made to yield the probability distribution. Consequently, this result in the examination of the error $E$. Over several repetitions of such an ensemble measurement, the updates of the angles are made until the measured $E$ becomes lower than the error tolerance $\epsilon_t$. We finally examine how many repetitions (iterations) of the ensemble measurements are required for the USFC to be optimal. This number is the key quantity in this work which determines the speed of machine learning. The setup described above also applies to cUSFC learning, but with an additional PBS inserted between the gates $g_0$ and $g_1$. The superposed state of the orthogonally polarized photons is collapsed to be either $H$- or $V$-polarized single-photon states, so that there is no quantum coherence involved. Apart from this, we apply the same scheme of measurement and feedback controller. Such a scheme for cUSFC learning leads to the same performance that would be obtained for the coherent state of light as the classical bit for the working channel.

\subsection{Learning algorithm} 

\begin{figure*}
\centering
\includegraphics[width=0.87\textwidth]{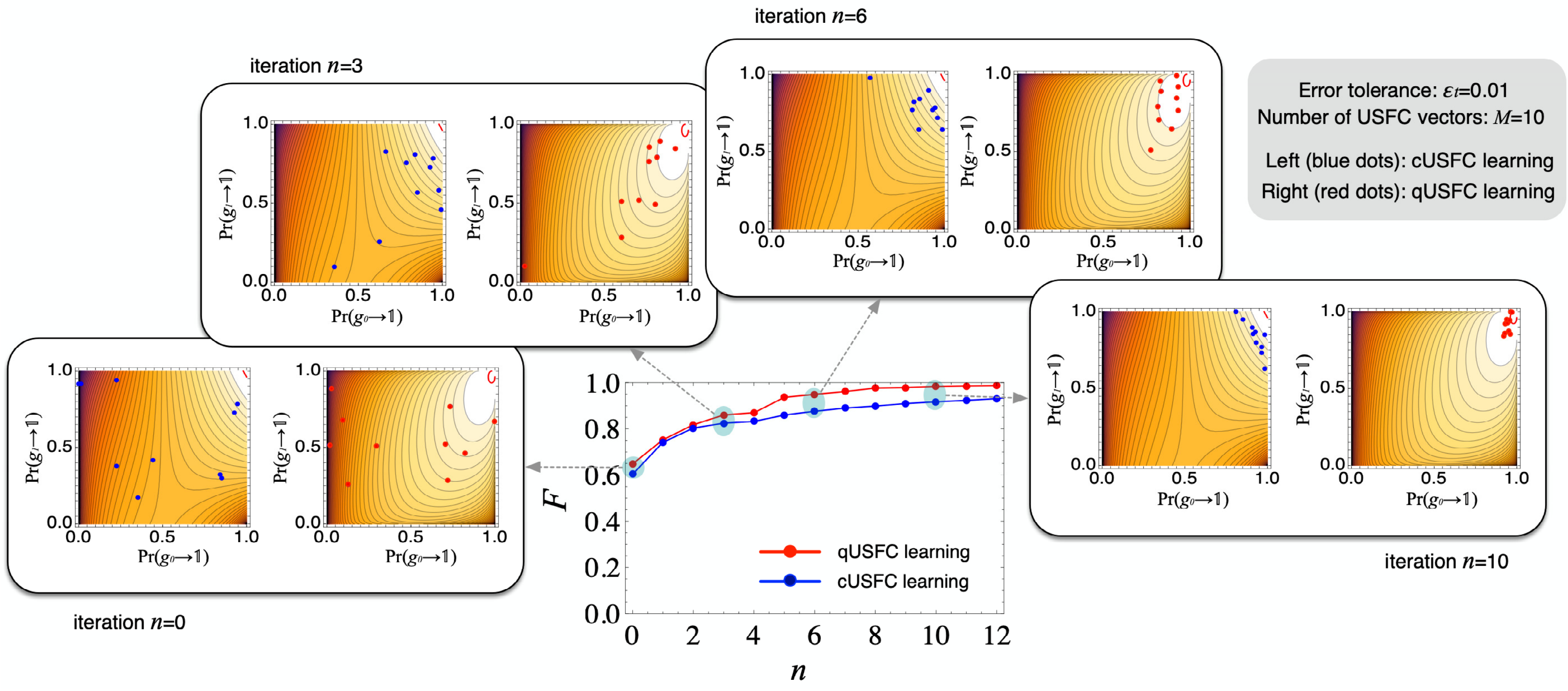}
\caption{\label{fig:DEprocess} {\bf Learning evolution of USFC.} Experimental task fidelity $F$ as a function of the iteration $n$. Contour plots show, for a selected trial, the evolution of the training USFC points in the parameter space of $\text{Pr}(g_0 \to \openone)$ and $\text{Pr}(g_1 \to \openone)$ as the learning (iteration) progresses. Here, $M=10$. The red line in the contour plot indicates the boundary of $F=0.99$ (or $\epsilon_t=0.01$).}
\end{figure*}

As the learning algorithm, we employ differential evolution (DE) which is known as one of the efficient global optimization methods~\cite{Storn97}. Our DE model considers $M$ agents, and thus $M$ gate-adapting preference vectors:
\begin{eqnarray}
\mathbf{p}_i = \Big(\text{Pr}(g_0 \to \openone), \text{Pr}(g_1 \to \openone) \Big)_i^\text{T}~(i=1,2,\ldots,M), 
\end{eqnarray}
each of which characterizes one USFC as a candidate, i.e., $\mathbf{p}_i \leftrightarrow \text{USFC}_i$. Here the task fidelity $F$ is used as a criterion that indicates how well the $\text{USFC}_i$ performs the target task (for the detailed method, see Appendix~B.1). In such settings, the vectors $\mathbf{p}_i$ are expected to evolve by `mating' their gate-adopting preferences. Therefore, the following aspect is commonly expected for both the classical and quantum cases; the USFC vectors $\mathbf{p}_i$ become closer together and move towards the exact solution point $\mathbf{p}_\text{sol}$. However, the theory predicts that the vectors $\mathbf{p}_i$ converge faster toward the solution point for qUSFC than cUSFC.

The experiments are performed using the DE algorithm. In the experiments, we assume the target is a constant function $f(x)=0$ ($x=0,1$). Here, $M=10$ and $\epsilon_t=0.01$ are chosen for both cUSFC and qUSFC. We set $\Delta=0$ in the case of qUSFC to maximize the contrast between cUSFC and qUSFC. Here, we define an experimental quantity $F$, such that 
\begin{eqnarray}
\widetilde{F} = \sqrt[4]{\left(\frac{L_{s}(0)}{L_\text{total}}\right)\left(\frac{L_{s}(1)}{L_\text{total}}\right)}, 
\end{eqnarray}
where $L_{s}(x)$ denotes the number of successful measurement outputs matched to the target for a given $x$, and $L_\text{total}$ denotes the total number of coincidence counts. We set $L_\text{total} = 10^5$ in our experiments. The obtained data are in a good agreement with our theoretical predictions and display the evolutionary behavior expected, as shown in Fig.~\ref{fig:DEprocess}. Here, note that the pre-settable DE parameters were chosen to maximize the efficiency of the cUSFC learning. This is understood as a penalty on the qUSFC. The advantage of qUSFC is better defined with this given handicap (for details, see Appendix~B.2).

\subsection{Learning probability} 

\begin{figure*}
\centering
\includegraphics[width=0.87\textwidth]{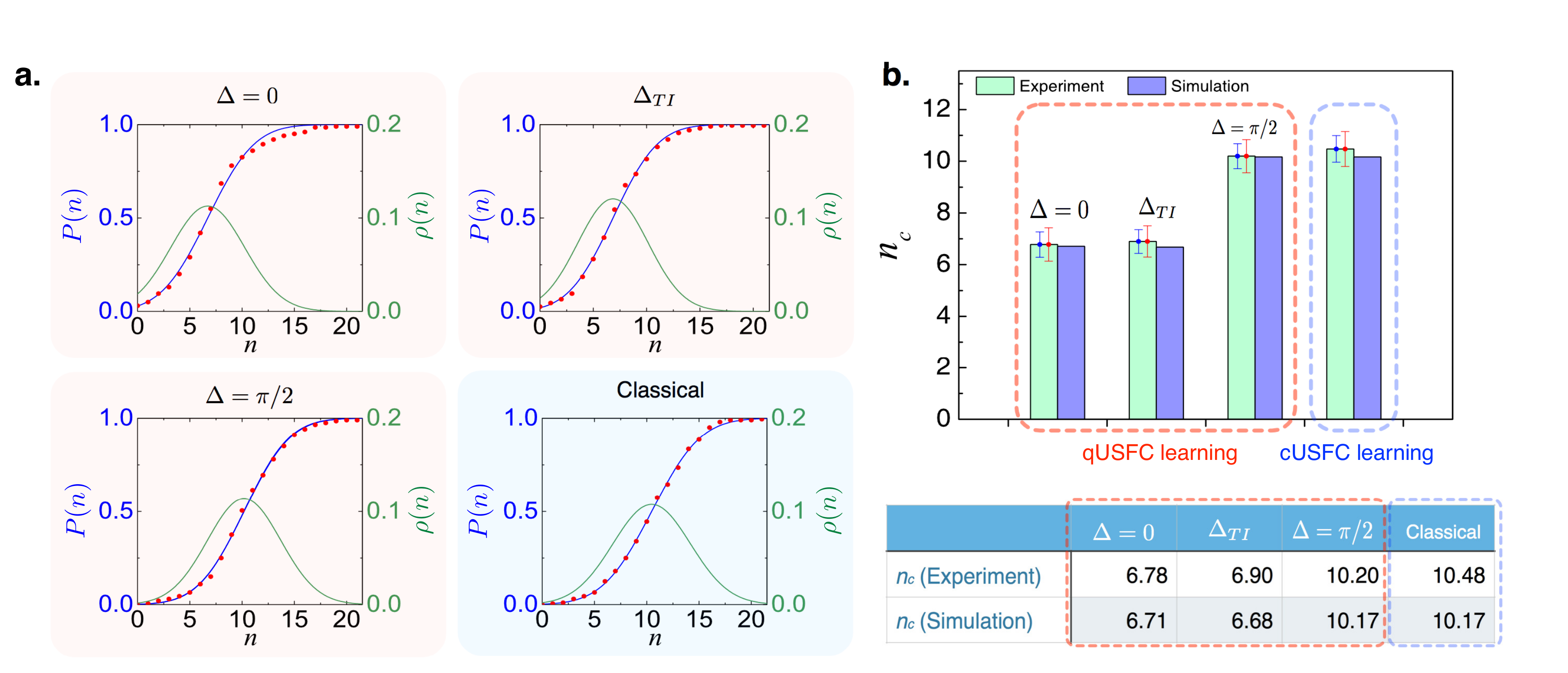}
\caption{\label{grp:result_qls} {\bf Experimental results for cUSFC and qUSFC learning.} {\bf a}. Here, the qUSFC learning experiments are performed with three $\Delta$-settings: $\Delta=0$, $\frac{\pi}{2}$, and $\Delta_{TI}$. The error tolerance $\epsilon_t$ is set to $0.01$. The data are created by counting the cases in which learning is completed before a certain iteration $n$, by sampling $200$ trials of the learning experiments. We present the learning probabilities $P(n)$ by fitting the counting data (red points). The data are well fitted (blue lines) by the integrated Gaussian $\int_{-\infty}^{n} dn' \rho(n')$, where $\rho(n) = \frac{1}{\sqrt{2\pi} \sigma_n} \exp\left({-\frac{(n-n_c)^2}{2\sigma_n}}\right)$. Here, we also depict $\rho(n)$ (green lines). {\bf b}. The average number of iteration $n_c$, the center position of $\rho(n)$, is compared for qUSFC and cUSFC. Here, the blue (red) bar represents the $95 (99) \%$ statistical confidence interval. Our results show quantum learning speed-up of $\simeq 36 \%$ for both $\Delta=0$ and $\Delta_{TI}$. All the experimental data are in good agreement with our analytical predictions.}
\end{figure*}

For statistical analysis, we repeat the experiments many times. The results are summarized in Fig.~\ref{grp:result_qls}. Firstly, we present the learning probability $P(n)$, which is defined as the probability of completing the learning process before a certain iteration $n$~\cite{Bang08}. The experiments are done for $M=10$ and $\epsilon_t=0.01$. The target is $f(x)=0$ ($x=0,1$). Here, qUSFC learning is performed for different $\Delta$ settings in order to maximize ($\Delta=0$) or eliminate ($\Delta=\frac{\pi}{2}$) the quantum advantages [see Eq.~(\ref{eq:F_cq})]. In addition, by noting that the appropriate choice of $\Delta$ depends on the target task, we consider a ``target-independent'' approach for the $\Delta$-setting, i.e., $\Delta_{TI} = \pi \Big( \text{Pr}(g_0 \to \openone) - \text{Pr}(g_1 \to \openone) \Big)$. There is no need for {\em a priori} knowledge about the target and a $\Delta$ setting is not required in advance. Choosing such a $\Delta_{TI}$ value makes the comparison between the classical and quantum cases as fair as possible, and more importantly, it is desirable for practical implementation. After the experiments are completed, $P(n)$ is characterized in each case with the average number of iterations, say $n_c$, required to complete learning (see Fig.~\ref{grp:result_qls}{\bf a}). We find that the qUSFC shows a faster learning convergence ($\simeq 36\%$ speed-up) as compared to cUSFC. Note that nearly the same degree of learning speed-up can be achieved for both $\Delta=0$ and $\Delta_{TI}$. We did not observe a speed-up for $\Delta = \frac{\pi}{2}$ (see Fig.~\ref{grp:result_qls}{\bf b}). For each case, we perform the learning trials $200$ times.

\subsection{Decoherence effect in qUSFC learning} 

\begin{figure}
\centering
\includegraphics[width=0.46\textwidth]{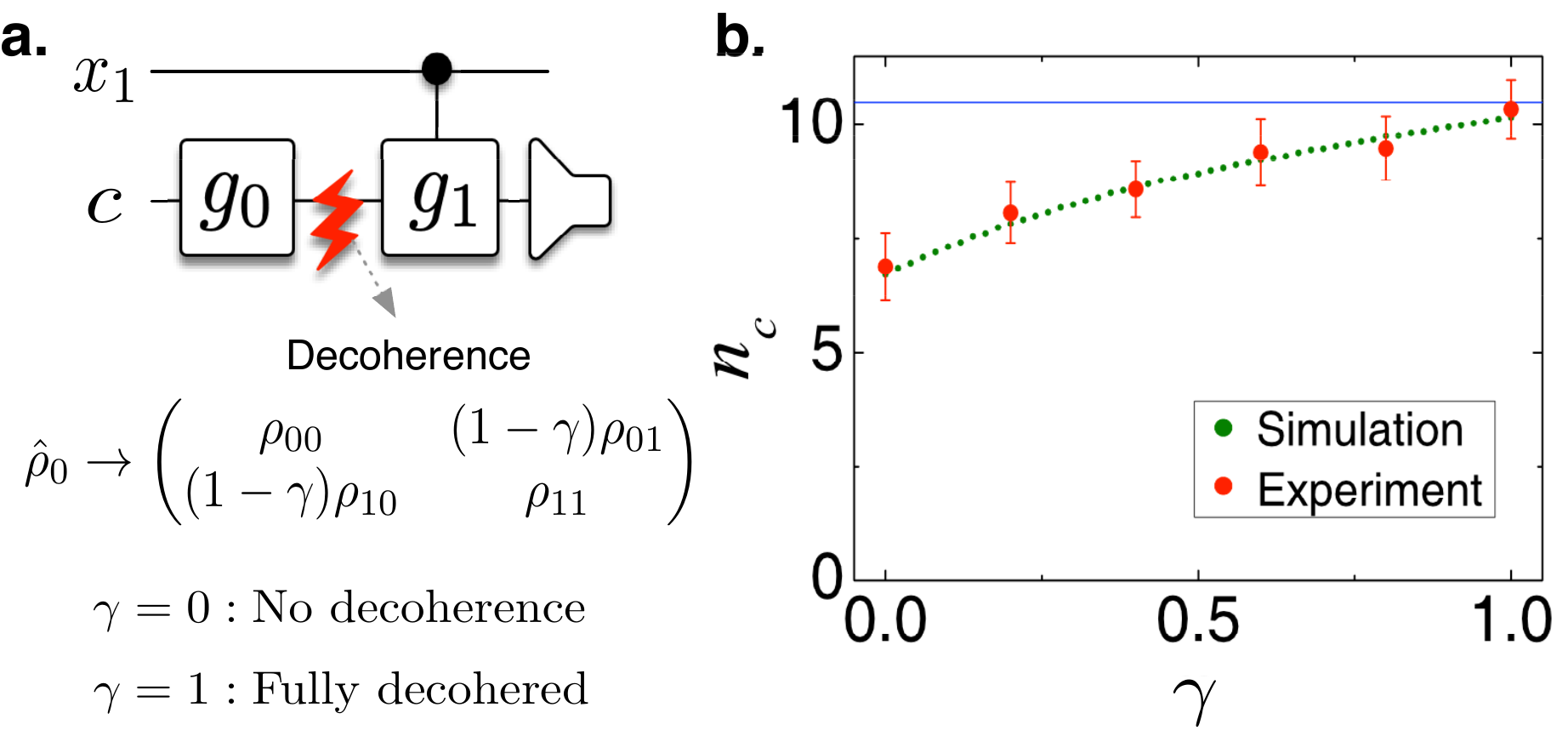}
\caption{\label{grp:result_decoherence} {\bf Decoherence effects in qUSFC learning.} {\bf a}. We assume that decoherence occurs between $g_0$ and $g_1$. {\bf b}. The experiments are performed for $\Delta=0$. The experimental values of $n_c$ (red solid circles) are shown together with numerically simulated data (green dotted). $100$ learning trials were used for each decay rate of $\gamma$. The blue line is for the classical approach.}
\end{figure}

We also investigate the effect of decoherence on qUSFC learning. For this purpose, we assume that the qubits passing through the gate operations decohere (decay of the off-diagonal elements in the density matrix, say, $\hat{\rho}_0$) at a rate of $\gamma \in [0,1]$ (see Fig.~\ref{grp:result_decoherence}{\bf a}). The decoherence results in the damage of the quantum advantage factor $\Lambda$ in Eq.~(\ref{eq:F_cq}), is given by:
\begin{eqnarray}
\Lambda \to \widetilde{\Lambda}=(1-\gamma)\Lambda. 
\end{eqnarray}
This implies that the qUSFC would not be able to {\em fully} exploit the quantum advantage. The experiments are performed for $\Delta=0$ by sampling $100$ learning trials for each decay rate of $\gamma$. The decoherence process is operated in our experiment by setting the relative phases of the states to either $0$ or as $\pi$ (a phase flip) randomly with a weight of $1-\gamma/2$ or $\gamma/2$. The results are shown in Fig.~\ref{grp:result_decoherence}{\bf b}, where the average number of iterations, $n_c$, increases as ($1-\gamma$) tends to $0$. However, the qUSFC maintains the learning speed-up unless the quantum superposition is completely destroyed. Note that the quantum learning speed-up persists by an arbitrary degree of the coherence. In the worst case, i.e., fully decohered case, the learning of qUSFC is identical to that of cUSFC.

\section{Remarks}

We proposed a quantum-classical hybrid scheme for machine learning in which the classical input data are unaltered and a small quantum working channel is employed. In particular, we employed the single-bit universal circuit, which can be extended to a large $N$-bit circuit of learning. We experimentally demonstrated quantum speed-up of approximately $36\%$ for learning using the single-bit universal circuit. We explained that the speed-up originates from quantum interference and the extension of the regions for approximate hypotheses; the extended region makes it easier to identify an approximate solution in the entire hypothesis space~\cite{Yoo14}. We showed that the quantum speed-up persists even in the presence of the dephasing noise before the quantum coherence is completely destroyed. The strong robustness against the dephasing noise is understood to be a remarkable feature of our hybrid machine learning. It is expected that this hybrid method will yield practical applications that exploit conventional classical techniques.

\section*{ACKNOWLEDGMENTS}
We thank M. Wie\ifmmode \acute{s}\else \'{s}\fi{}niak, W. Laskowski, M. Paw\l{}owski, N. Liu, and J. Fitzsimons for helpful discussions. We acknowledge the financial support of the Basic Science Research Program through the National Research Foundation of Korea (NRF) grant (No. 2014R1A2A1A10050117 and No. 2016R1A2B4014370) funded by the Ministry of Science, ICT \& Future Planning. JB also acknowledges the financial support of the ICT R\&D program of MSIP/IITP (No. 10043464).

\appendix

\section{Primitive strategy for general $N$-bit data classification}

\begin{figure}
\centering
\includegraphics[width=0.46\textwidth]{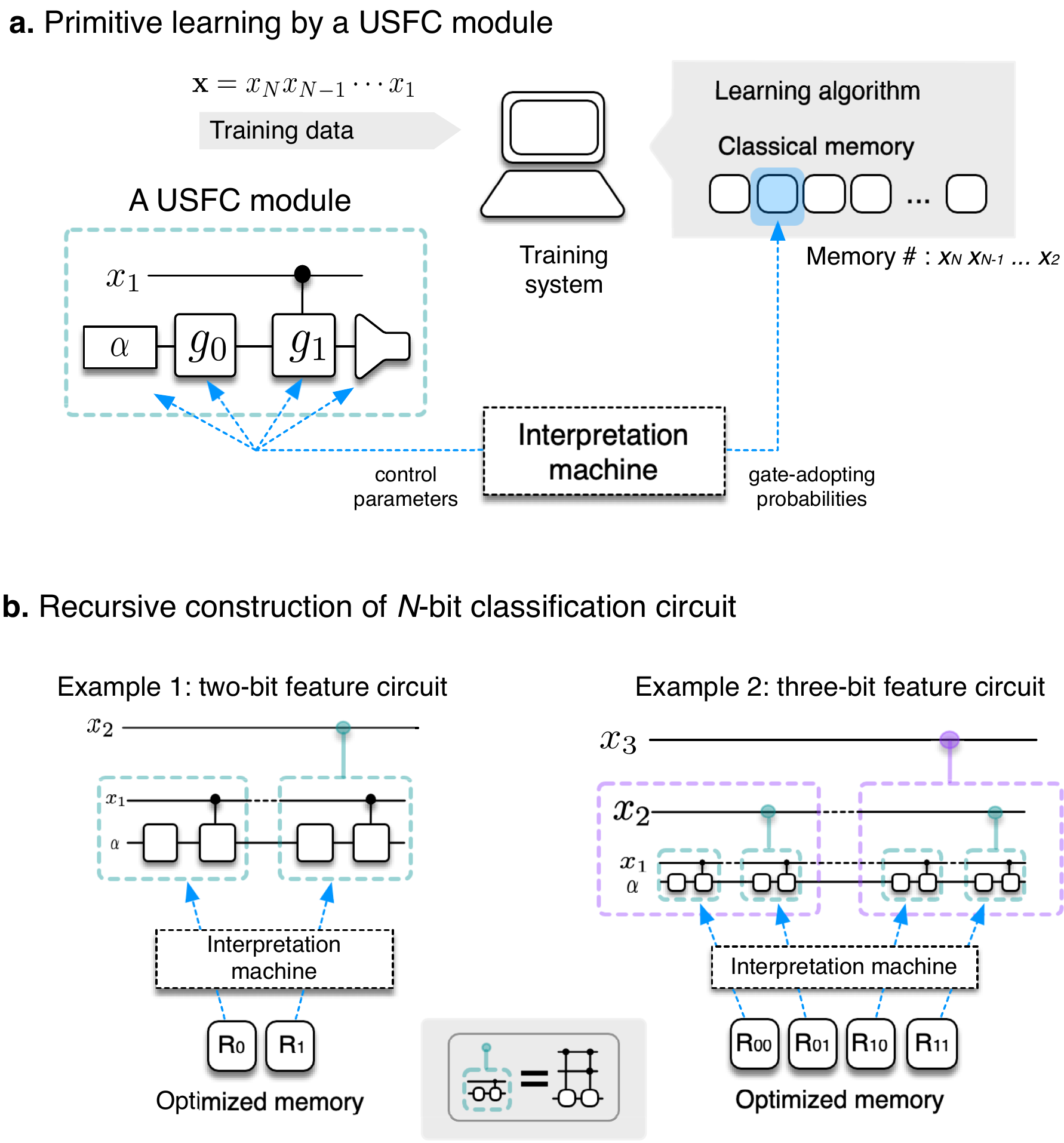}
\caption{\label{fig:primitive_learning} Schematic view of the primitive strategy for larger $N$. {\bf a}. Basic structure: a single USFC and training system are casted. The training system includes the learning algorithm and $2^{N-1}$ memory blocks. {\bf b}. After completion of learning, the entire N-bit classification circuit can be constructed.}
\end{figure}

Here, we introduce a naive strategy for $N$-bit binary classification learning, which consists of a USFC module and a training system. The USFC module is equipped with controllable preparation and measurement devices. The training system includes the learning algorithm (e.g., differential evolution in our case) and $2^{N-1}$ classical memory blocks, each of which stores the corresponding gate-adopting preferences (see Fig.~\ref{fig:primitive_learning}{\bf a}). In such a setting, the learning proceeds as follows. Firstly, a supervisor (or a server) provides a set of training inputs $\mathbf{x} = x_N x_{N-1} \cdots x_1$ and targets $T_\mathbf{x} \in \{0,1\}$. The training system initializes and arranges the gate-adopting preferences in the classical memory blocks. Then, by iterating the following [{\bf A.1}]-[{\bf A.2}], the USFC module optimizes all $2^{N-1}$ sets of the gate-adopting preferences.
\begin{description}
	\item[{[A.1]}] --- For an input $\mathbf{x} = x_N x_{N-1} \cdots x_1$ entering the USFC, the training system calls the gate-adopting preferences $\Big( \text{Pr}(g_0 \to \openone), \text{Pr}(g_1 \to \openone) \Big)_\mathbf{j}$ stored in the $\mathbf{j}$-th memory block, say $R_\mathbf{j}$. Here, the memory index $\mathbf{j}$ is identified by the data bits of the input features $x_N x_{N-1} \cdots x_2$, except for the last one $x_1$, i.e., $\mathbf{j}=x_N x_{N-1} \ldots x_2$.
	
	\item[{[A.2]}] --- Then, USFC learning is performed with the single-feature input $x_1$. In this learning process, the input $\alpha$ and the measurement target $\tau$ of the working channel are determined according to:
	\begin{eqnarray}
	\alpha = 
	\left\{
	\begin{array}{ll}
 	0 & \text{if}~\mathbf{j} = 00 \ldots 0, \\
	T_{\mathbf{j}\oplus(-0 \ldots 01),x_1} \in \{0,1\} & \text{if}~\mathbf{j} \neq 00 \ldots 0,
	\end{array}
	\right.
	\end{eqnarray}
	and $\tau = T_{\mathbf{j},x_1} \in \{0,1\}$.
\end{description}
After completion of learning, the entire $N$-bit classification circuit can be constructed, recursively, with $2^{N-1}$ different USFC modules, each of which is obtained based on the optimized gate-adopting preferences in $R_\mathbf{j}$. 

As a simple example, let us consider the two-bit feature ($N=2$) binary classification learning, where the training input $\mathbf{x} = x_2 x_1 \in \{ 00, 01, 10, 11 \}$ and target $T_{x_2 x_1} \in \{0, 1\}$ are provided. In this case, the training system uses two classical memory blocks $R_\mathbf{j}$ (here, $\mathbf{j}=x_2 \in \{0,1\}$). USFC learning is performed with the last bit $x_1$ of the input $\mathbf{x}$. The working input $\alpha$ and the single-bit target $\tau$ are selected using [A.2], i.e.,
\begin{eqnarray}
\text{if}~x_2=0
\left\{
\begin{array}{l}
\alpha=0, \\
\tau=T_{0 x_1}, \\
\end{array}
\right. ~\text{and,}~
\text{if}~x_2=1
\left\{
\begin{array}{l}
\alpha=T_{0 x_1}, \\
\tau=T_{1 x_1}. \\
\end{array}
\right.
\end{eqnarray}
After learning is completed for all the memory blocks, the two-bit classification circuit can be constructed with the optimized gate-adopting preferences in $R_0$ and $R_1$, as shown in Fig.~\ref{fig:primitive_learning}{\bf b}. One can see how such a primitive strategy works for higher $N$-bit training data. 

Now, we indicate that the quantum learning speed-up of the qUSFC described in this work can be generalized to the case of larger $N$ by means of the primitive learning strategy described in the preceding section. Roughly, the overall learning time would be $2^{N-1} T_C$ and $2^{N-1} T_Q$ for an arbitrary $N$, where $T_C$ and $T_Q$ are the learning time for the cUSFC and qUSFC, respectively (here, $T_C > T_Q$). We clarify here that achieving quantum learning speed-up in this manner would be the worst-case scenario. In other words, {\em the quantum learning speed-up analyzed by a single USFC would be the lower bound for arbitrary $N$-bit binary classification learning.} Actually, such a speed-up is attributable to the quantum coherence of the two different unitary gates involved only in a single qUSFC. If the coherence of all $2^N$ unitary gates is used without any intermediate measurements between the different qUSFCs, the degree of quantum learning speed-up will be even more pronounced (for details, see our previous theoretical results in Ref.~\cite{Yoo14}).

\section{Differential evolution}

\subsection{Method} 

One of the most important aspects of machine learning is the choice of the learning algorithm since the efficiency of learning strongly depends on the algorithm. In this work, we choose differential evolution (DE), which is one of the most efficient global optimization methods~\cite{Storn97}. By employing the DE algorithm, we first prepare $M$ gate-adopting preference vectors as candidates: $\mathbf{p}_i = (p_0, p_1)^T$ for $i=1,2,\cdots,M$, where $p_0$ and $p_1$ denote $\text{Pr}(g_0 \to \openone)$ and $\text{Pr}(g_1 \to \openone)$, respectively. They are chosen initially at random and saved in the classical memory of the training system. We then implement differential evolution as follows:

\begin{description}
	\item[Step 1.] --- For each $\text{USFC}_i$, we generate $M$ mutant vectors $\boldsymbol\nu_{i} = \left(\nu_0, \nu_1\right)_i^T$, according to 
\begin{eqnarray}
\boldsymbol\nu_i = \mathbf{p}_a + W \left(\mathbf{p}_b - \mathbf{p}_c\right),
\end{eqnarray}
where $\mathbf{p}_a$, $\mathbf{p}_b$, and $\mathbf{p}_c$ are randomly chosen for $a,b,c \in \{1,2,\cdots,M\}$. These three vectors are chosen to be different from each other; hence, it is necessary that $M \ge 3$. The free parameter $W$, which is called the differential weight, is a real and constant number. 

	\item[Step 2.] --- Subsequently, all $M$ parameter vectors $\mathbf{p}_i = \left( p_0, p_1 \right)_i^T$ are reformed to trial vectors $\mathbf{t}_{i} = \left( t_0, t_1 \right)_i^T$ by the following rule: For each $j \in \left\{ 0, 1 \right\}$,
\begin{eqnarray}
\label{eq:crossover}
\left\{
\begin{array}{ll}
t_{j} \leftarrow p_{j} & ~~\textrm{if}~r_j > C_r,\\
t_{j} \leftarrow \nu_{j} & ~~\textrm{otherwise}, \\
\end{array}
\right.
\end{eqnarray}
where $r_j \in [0, 1]$ is a random number and the crossover rate $C_r$ is another free parameter which lies between $0$ and $1$. Note that $W$ and $C_r$ are set to achieve the best learning efficiency. 

	\item[Step 3.] --- Finally, $\mathbf{t}_i$ is taken for the next iteration if the newly updated $\text{USFC}_i$ from $\mathbf{t}_i$ yields a higher fitness value than the previous one from $\mathbf{p}_i$; if not, $\mathbf{p}_i$ is retained. Here, the fitness is defined as the task fidelity $F$. While evaluating the $M$ fitness values, the training system records the best $F_\text{best}$ and $\mathbf{p}_\text{best}$ in the classical memory. 
\end{description}

Steps $1$-$3$ described above are repeated until $F_\text{best}$ is maximized. Ideally, our DE algorithm is supposed to find $\mathbf{p}_\text{best}$ that yields $F_\text{best} \simeq 1 - \epsilon_t$.

\subsection{Setting of the free parameters $W$ and $C_r$} 

\begin{figure*}
\centering
\includegraphics[width=0.86\textwidth]{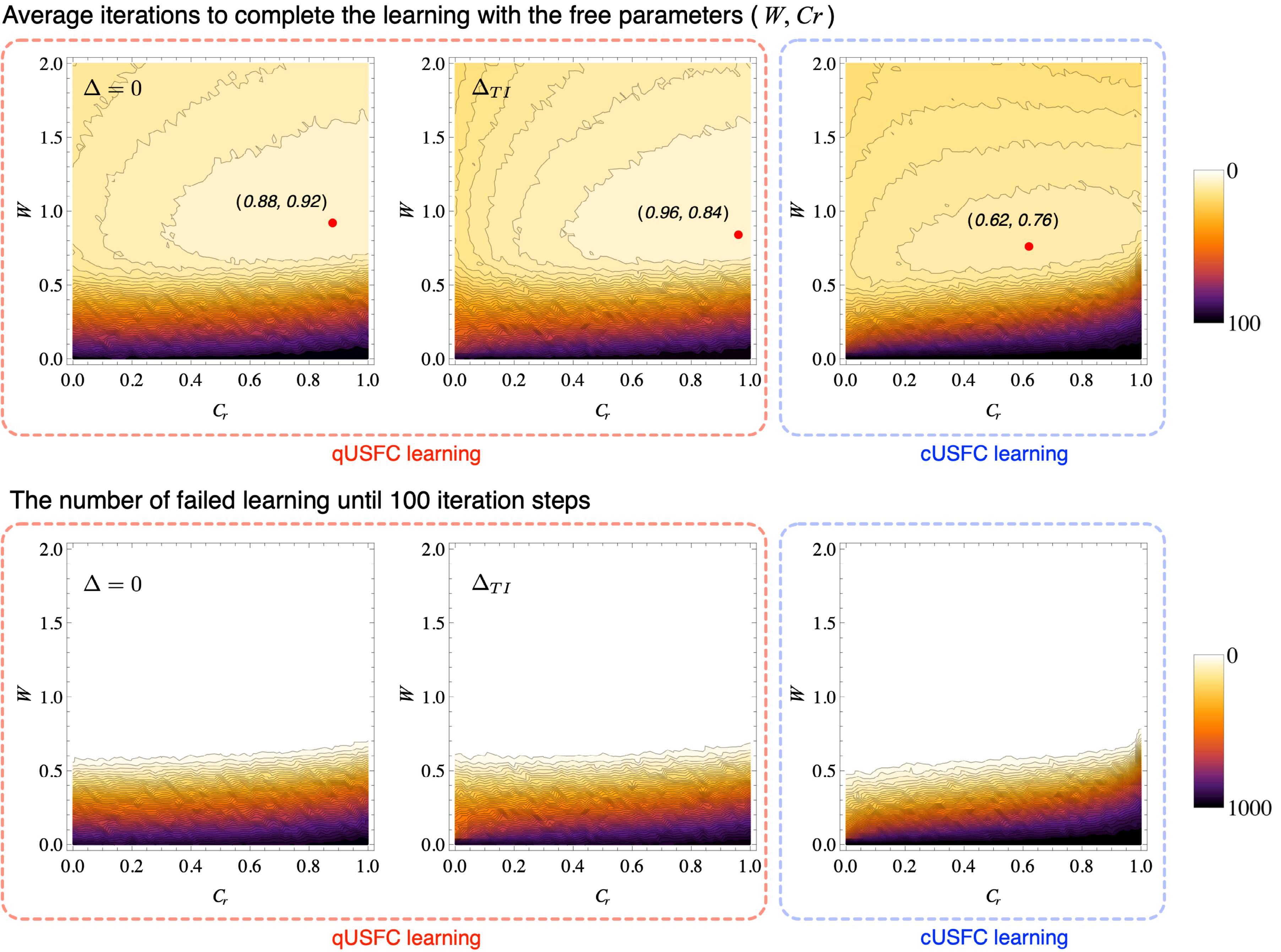}
\caption{\label{fig:set_f_param} By performing numerical simulations, we determined the average number of iterations, $n_c$, by varying the free parameters $W$ and $C_r$, and obtained the density plot of $n_c$ in the two-dimensional space of ($C_r$, $W$). The qUSFC simulations are performed for $\Delta = 0$, $\frac{\pi}{2}$, and $\Delta_{TI}$. Here, the best-optimized parameters ($C_{r,\text{best}}$, $W_\text{best}$) are found in each case (depicted by the red points). We also obtained the density plots of the number of failed learning processes out of $1000$ trials. For failed learning the task fidelity could not reach $0.99$ until $100$ evolution steps, which indicates that the chosen parameter values are practically unavailable.}
\end{figure*}

In most machine learning algorithms, it is very important to set the learning parameters because the learning efficiency strongly depends on the chosen parameter values. In our case, the parameters $W$ and $C_r$ are chosen to maximize the learning efficiency and we are free to use pre-established data or empirical knowledge without any specific rules~\cite{Storn97}. Thus, in this case, we construct a priori data, which are used to set the free parameter values $W$ and $C_r$ by performing the numerical simulations. The simulations are performed for cUSFC and qUSFC by varying $W$ (from $0$ to $2$) and $C_r$ (from $0$ to $1$), with $1000$ learning trials. In the case of qUSFC learning, we consider three different $\Delta$ settings: $\Delta=0$, $\Delta=\frac{\pi}{2}$, and $\Delta_{TI}=\pi \Big( \text{Pr}(g_0 \to \openone) - \text{Pr}(g_1 \to \openone) \Big)$. Figure~\ref{fig:set_f_param} shows the results, where the average iteration number $n_c$ required to complete learning is depicted as density plots in the plane of ($W$, $C_r$), and the optimized parameter values that exhibit the best learning efficiency are found in each case. The number of instances of failed (i.e., unconverged) learning up to $100$ evolution steps is also determined. By observing these results, we choose $W$ and $C_r$ that optimize the cUSFC learning (not of qUSFC learning). Such a choice might be considered as a penalty for qUSFC learning, because the learning efficiency strongly depends on the parameters of the algorithm. Nevertheless, qUSFC learning is shown to be faster and thus overcomes this limitation (see our main manuscript).

\bibliographystyle{aps}

\end{document}